\documentclass[usenatbib,usegraphicx]{mn2e}
\usepackage{subfigure}
\usepackage{longtable}
\usepackage{graphicx}
\usepackage{graphics}
\usepackage{keyval}
\usepackage{trig}
\def\beq{\begin{equation}}
\def\eeq{\end{equation}}
\def\bey{\begin{eqnarray}}
\def\eey{\end{eqnarray}}

\def\msun{M_\odot}

\def\lsim{\mathrel{\raise.3ex\hbox{$<$\kern-.75em\lower1ex\hbox{$\sim$}}}}
\def\gsim{\mathrel{\raise.3ex\hbox{$  $\kern-.75em\lower1ex\hbox{$\sim$}}}}

\def\kms{\, {\rm km \, s}^{-1} }

\def\grad{{\bf \nabla}}

\title{The collision velocity of the bullet cluster in conventional and modified dynamics}
\author[G. W. Angus and S. S. McGaugh]{G. W. Angus$^{1}$\thanks{email:
gwa2@st-andrews.ac.uk}, S. S. McGaugh$^{2}$\thanks{email:
ssm@astro.umd.edu} \\
$^{1}$SUPA, School of Physics and Astronomy, University of St. Andrews, Scotland KY16 9SS\\
$^{2}$Department of Astronomy, University of Maryland, College Park, MD 20742-242 USA\\}

\begin{document}

\date{Accepted ... Received ... ; in original form ...}

\pagerange{\pageref{firstpage}--\pageref{lastpage}} \pubyear{2007}

\maketitle

\label{firstpage}
\begin{abstract}
We consider the orbit of the bullet cluster 1E 0657-56 in both CDM and MOND using accurate mass models appropriate to each case in order to ascertain the maximum plausible collision velocity. Impact velocities consistent with the shock velocity ($\sim 4700\kms$) occur naturally in MOND. CDM can generate collision velocities of at most $\sim 3800\kms$, and is only consistent with the data provided that the shock velocity has been substantially enhanced by hydrodynamical effects.
\end{abstract}

\begin{keywords}
gravitation - dark matter - galaxies: clusters: individual (1E 0657-56)
\end{keywords}

\section{Introduction}
\protect\label{sec:intr}

Many lines of observational evidence now oblige us to believe that the universe is filled with a novel, invisible form of mass that dominates gravitationally over normal baryonic matter.  In addition, a dark
energy component which exerts negative pressure to accelerate the
expansion of the universe is also necessary (Chernin et al. 2007).  Though this $\Lambda$CDM
paradigm is well established, we still have only ideas about what these
dark components might be, and no laboratory detections thereof.

One possible alternative to $\Lambda$CDM is the 
Modified Newtonian Dynamics (MOND; Milgrom 1983a,b,c).
This hypothesis has been more successful than seems to be widely
appreciated (McGaugh \& de Blok 1998; Sanders \& McGaugh 2002),
and has received a theoretical boost from the introduction of generally
covariant formulations (Bekenstein 2004; Sanders 2005; Zlosnik, Ferreira, \& Starkman 2006,7).  The dark matter and alternative gravity paradigms are radically different, so every observation that might distinguish between them is valuable.

\begin{figure}
\includegraphics[angle=0,width=8.5cm]{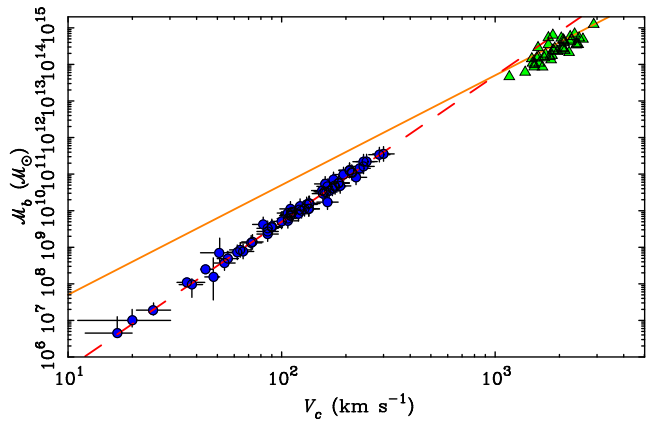}
\caption{Shows baryonic mass against circular velocity. Rotating galaxies (blue circles) are from McGaugh (2005) and clusters (green triangles) are from Sanders (2003) using the measured temperature to estimate the circular velocity assuming isothermality. The solid orange line is the CDM M-V relation (Steinmetz \& Navarro 1999) assuming $M_b = f_b M_{vir}$ with $f_b=0.17$ (Spergel et al.\ 2006) and the dashed red line is the MOND prediction. The spirals lie directly on the MOND prediction, but the clusters are generally 2-3 times in mass below it.  The CDM expectation is nicely consistent with clusters, but implies many dark baryons in spirals in addition to the non-baryonic dark matter.}
\protect\label{fig:btf}
\end{figure}

$\Lambda$CDM is known to work well on large scales (e.g., Spergel et al.\ 2006) while MOND is known to work well in individual galaxies (Sanders \& McGaugh 2002).  This success, incorporating the tight correlation between dark and luminous mass in the DM framework (McGaugh 2005; Famaey et al. 2007b) extends over five decades in mass (Fig.\ref{fig:btf}) ranging from tiny dwarfs (e.g., Milgrom \& Sanders 2007) through spirals of low surface brightness (de Blok \& McGaugh 1998), our own Milky Way (Famaey \& Binney 2005) and other high surface brightness (Sanders 1996; Sanders \& Noordermeer 2007) to massive ellipticals (Milgrom \& Sanders 2003). The recent observations of tidal dwarf galaxies by Bournaud et al. (2007) provides a severe challenge to CDM but is naturally explained in MOND with {\bf zero} free parameters (Milgrom 2007; Gentile et al. 2007). Having said that, MOND persistently fails to completely explain the mass discrepancy in rich clusters of galaxies.  Consequently, clusters require substantial amounts of non-luminous matter in MOND. 

That rich clusters contain more mass than meets the eye in MOND goes back to Milgrom's original papers (Milgrom 1983c).  At the time, the discrepancy was very much larger than it is today, as it was not then widely appreciated how much baryonic mass resides in the intra-cluster medium.  Further work on the X-ray gas (e.g., Sanders 1994, 1999) and with velocity dispersions (McGaugh \& de Blok 1998) showed that MOND was at least within a factor of a few, but close inspection revealed a persistent discrepancy of a factor of two or three in mass (e.g., Gerbal et al.\ 1992; The \& White 1998; Pointecouteau \& Silk 2005, Buote \& Canizares 1994).  Weak gravitational lensing (Angus et al.\ 2007a; Takahashi \& Chiba 2007; Famaey, Angus et al. 2007) provides a similar result.

To make matters worse, the distribution of the unseen mass does not trace that of either the galaxies or the X-ray gas (Aguirre et al.\ 2001; Sanders 2003; Angus et al. 2007b; Sanders 2007). In Fig. \ref{fig:btf} we plot the baryonic mass of many spiral galaxies and clusters against their circular velocity together with the predictions of MOND and CDM. MOND is missing mass at the cluster scale.  CDM suffers an analogous missing baryon problem on the scale of individual galaxies.

The colliding bullet cluster 1E-0657-56 (Clowe et al. 2004,2006, Bradac et al. 2006, Markevitch et al. 2004, Markevitch \& Vikhlinin 2007) illustrates in a spectacular way the residual mass discrepancy in MOND.  While certainly problematic for MOND as a theory, it does not constitute a falsification thereof.  Indeed, given that the need for extra mass in clusters was already well established, it would have been surprising had this effect not also manifested itself in the bullet cluster.  The new information the bullet cluster provides is that the additional mass must be in some collision-less form.

It is a logical fallacy to conclude that because extra mass is required by MOND in clusters, that dark matter is required throughout the entire universe.  While undeniably problematic, the residual mass discrepancy in MOND is limited to groups and rich clusters of galaxies: these are the only systems in which it systematically fails to remedy the dynamical mass discrepancy (see discussion in Sanders 2003).  Could we be absolutely certain that we had accounted for all the baryons in clusters, then MOND would indeed be falsified.  But CDM suffers an analogous missing baryon problem in galaxies (Fig.\ref{fig:btf}) in addition to the usual dynamical mass discrepancy, yet this is not widely perceived to be problematic.  In either case we are obliged to invoke the existence of some dark mass which is presumably baryonic (or perhaps neutrinos) in the case of MOND.  In neither case is there any danger of violating big bang nucleosynthesis constraints.  The integrated baryonic mass density of rich clusters is much less than that of all baryons; having the required mass of baryons in clusters would be the proverbial drop in the bucket with regards to the global missing baryon problem.

A pressing question is the apparently high relative velocity between the two clusters that comprise the bullet cluster 1E 0657-56 (Clowe et al. 2006, Bradac et al. 2006, Markevitch et al. 2004, Markevitch \& Vikhlinin 2007).  The relative velocity derived from the gas shockwave is $v_{rel}=4740_{-550}^{+710}\kms$ (Clowe et al.\ 2006). Taken at face value, this is very high, and seems difficult to reconcile with $\Lambda$CDM (Hyashi \& White 2006).  The problem is sufficiently large that it has been used to argue for an additional long range force in the dark sector (Farrar \& Rosen 2007).  Here we examine the possibility of such a large velocity in both CDM and MOND.

One critical point that has only very recently been addressed is how the shock velocity relates to the collision velocity of the clusters.  Naively, one might expect the dissipational collision of the gas clouds to slow things down so that the shock speed would provide a lower limit on the collision speed.  Recent hydrodynamical simulations (Springel \& Farrar 2007; Milosavljevic et al.\ 2007) suggest the opposite.  A combination of effects in the two hydrodynamical simulations show that the shock velocity may be higher than the impact velocity.  The results of the two independent hydrodynamical simulations do not seem to be in perfect concordance, and the precise result seems to be rather model dependent.  Nevertheless, it seems that the actual relative velocity lies somewhere in the range 3500-4500$\kms$.

The difficulties posed by a high collision velocity for CDM have been discussed previously by Hayashi \& White (2006) and Farrar \& Rosen (2007).  Whereas Springel \& Farrar (2007) and Milosavljevic et al. (2007) consider the complex hydrodynamic response of the two gas clouds during the ongoing collision, here we investigate the ability of two clusters like those comprising the bullet cluster to accelerate to such a high relative velocity in the case of both CDM and MOND prior to the merger.  We compute a simple free fall model for the two clusters in an expanding universe with realistic mass models, and ask whether the observed collision velocity can be generated within the time available. We take care to match the mass models to the specific observed properties of the system appropriate to each flavor of gravity in order to realistically evaluate the orbit of the clusters prior to their collision.

\section{Modeling the freefall}

We wish to address a simple question.  Given the observed masses of the two clusters, is it possible to account for the measured relative velocity from their gravitational freefall?  The expansion of the universe mitigates against large velocities, since the clusters must decouple from the Hubble flow before falling together.  Presumably it takes some time to form such massive objects, though this is expected to occur earlier in MOND than in $\Lambda$CDM (Sanders 1998, 2001; McGaugh 1999, 2004; Nusser 2002; Stachniewicz \& Kutschera 2002; Knebe \& Gibson 2004; Dodelson \& Liguori 2006).  The clusters are observed at z=0.3, giving at most 9Gyrs for them to accelerate towards each other.  This imposes an upper limit to the velocity that can be generated gravitationally.  Without doing the calculation, it is not obvious whether the larger masses of the clusters in CDM or the stronger long range force in MOND will induce larger relative velocities. 

Since we know the state of the system directly prior to collision, it makes sense to begin our simulations from the final state and work backwards in time towards when the relative velocity was zero. This point, where the clusters have zero relative velocity, is when they turned around from the Hubble flow and began their long journey gravitating towards each other. Working backwards in time leads to potentially counter-intuitive discussions (such as Hubble contraction), which we try to limit. 

We must account for the Hubble expansion in a manner representing the universe before z=0.3.  The detailed form of the expansion history of the universe $a(t)$ is not known in the case of MOND, so we take the scale factor of $\Lambda CDM$ in both models

\beq
\label{eqn:sf}
{da(t) \over dt}=H_o\left[ \Omega_m a^{-1} +\Omega_{\Lambda} a^2    \right]^{1/2}.
\eeq
Where we take $H_o=72 \kms Mpc^{-1}$, $\Omega_m$=0.27 and $\Omega_{\Lambda}=0.73$.

The important aspect is the basic fact that the universe is expanding and the mutual attraction of the clusters must overcome this before they can plunge together at high velocity.  

We implement the scale factor in the simulations through the equation of motion
\beq
\label{eqn:eom}
{1 \over a(t)}{d \over dt}  [a(t)v]=g.
\eeq
Computing this numerically, from time step to time step we calculate the ratio of the scale factor in the previous time step to the current time step (i.e. $\epsilon=a(t_{i-1})/a(t_{i})$; we use negative time steps to move backwards in time from the presently known configuration, so $\epsilon>1$ (higher $i$ means earlier universe). We then have $v(t_{i})=v(t_{i-1})\epsilon+g \Delta t$.

The right hand side of Eq. \ref{eqn:eom} differs in MOND and CDM not only because the law of gravity is altered, but also because the gravitating masses are higher in CDM.

The initial conditions are the crux of the problem, with at least 4 unknowns.  These include the masses of the two clusters, the relative velocity of the clusters, and the distance of separation between the two when they had this relative velocity. The separation is the same in MOND and Newtonian gravity, but the Newtonian mass is higher.  

The relative velocity of the two clusters can be measured because, in the last few 100Myrs, the less massive sub cluster has passed through the centre of the more massive main cluster.  The ram pressure has imposed a smooth bow shock (Markevitch et al. 2004, Markevitch \& Vikhlinin 2007) on the gas of the sub cluster. Since the relative velocity is the foundation of the problem we leave it free and try to estimate it by fixing other variables. In our simulation, we think it sensible to consider the separation of the two clusters (i.e. of the two centres of mass) when they had the calculated relative velocity to be when the leading edge of the sub cluster's gas cloud began to pass through the dense region of gas belonging to the main cluster and separate from the dark matter. It appears that the centre of the sub cluster's gas cloud (the location of the bullet) is preceeded by the bow shock by around 200kpc further in the direction of travel. We take 200kpc to also be the radius within which the gas of cluster 1 was dense enough to imprint the bow shock. Indeed, the gas mass of the main cluster (sub cluster) is only measured out to 180kpc (100kpc) and could not be found further detailed in the literature. However, the uncertainty is large enough that we wished to clarify the impact of different initial separations by always using a range of initial separations of between 350-500kpc. This separation is defined as when the two pre-collision clusters had the relative velocity of $v_{rel}$ related to the shock velocity $4740_{-550}^{+710}\kms$. Now of course, they are on the opposite sides on the sky after having passed through each other and the gas has been offset from the DM. 

\begin{figure}
\includegraphics[angle=0,width=8.5cm]{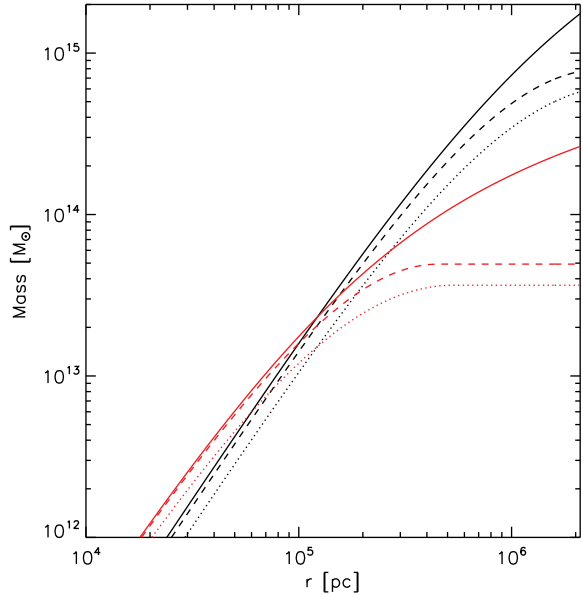}
\caption{The total enclosed masses for the main cluster (black) and sub cluster (red) for CDM (solid), MOND with standard $\mu$ (dashed) and simple $\mu$ (dotted). }
\protect\label{fig:masses}
\end{figure}

\begin{figure}
\includegraphics[angle=0,width=8.5cm]{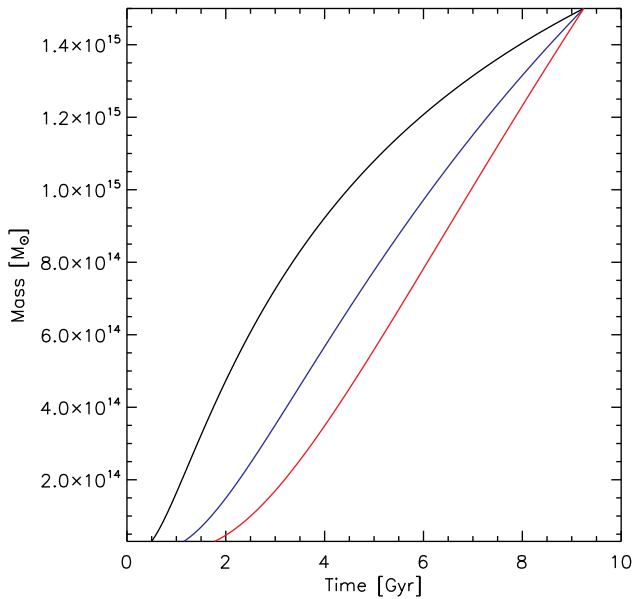}
\caption{The total enclosed masses for the main cluster as a function of time, where 9.3Gyr marks the collision of the two clusters and 0Gyr represents the Big Bang. The three lines correspong to different mass assembly rates $\alpha$=0.5 (black), 1.0 (blue) and 1.5 (red). The mass loss is halted at $m_{200}$/50.}
\protect\label{fig:mz}
\end{figure}

\section{The collision in CDM}
\protect\label{sec:newtbul}

In the CDM framework, it is no problem to generate two clusters in an N-body simulation and calculate all gravitational accelerations exactly. However, in MOND we are dealing with non linear gravity and the tools for such purposes are only now being developed (Nipoti et al. 2007a). Additionally, since we begin our simulations with the overlap of the two clusters, it is not guaranteed that the clusters preserve their shapes as they separate. Furthermore, it was not possible to simply include accretion history (Wechsler et al. 2002) in the N-body simulations or easily vary the truncation radius of the DM halo as was necessary.

Therefore, a better method was to semi-analytically account for these aspects in a simulation where gravity of one cluster acting on the other is just the mass enclosed by a sphere around the gravitating body's centre of mass with radius equal to the separation of the two cluster's centres of mass. For this procedure, the only two unknowns are the separation, which is initially known and computed each time-step; and the mass enclosed. The enclosed mass depends on the density profile of the two clusters and was fitted by Clowe et al. (2006) using NFW profiles of the form (see Angus \& Zhao 2007)

\begin{equation}
\rho(r) = {A \over 4\pi} m_{200} r^{-1} \left(r+r_{200} c^{-1}\right)^{-2}, {\rm } A^{-1}=\ln(1+c) - {c \over 1+c}
\end{equation}
where c is the concentration. The enclosed mass goes as
\begin{equation}
\protect\label{eqn:nfwmass}
m(r) = A m_{200} \left[ \ln \left(1 + {cr \over r_{200}} \right) - 1 + \left(1 + {cr \over r_v}\right)^{-1}  \right] 
\end{equation}

For the main cluster they give $m_{200}=1.5 \times 10^{15}\msun$, $r_{200}=2100kpc$ and concentration c=1.94. For the sub cluster, $m_{200}=1.5 \times 10^{14}\msun$, $r_{200}=1000kpc$ and c=7.12. We augment the DM with a baryon fraction of 17\% (Spergel et al.\ 2006) which was part of the total mass during freefall. We ran dynamical time steps (negative) such that

\beq
\Delta t=10^{-4} \times \sqrt{d[pc]}Myr
\eeq

Where d is the separation of the two centres of mass and $\Delta t$ has a maximum value of 1Myr. Since initially d $\sim$ 400kpc, the starting time steps are $\sim$0.06Myr. The simulations were run until 9Gyr had elapsed.

The mass distributions as functions of radius for the two clusters in CDM and the ones used in the MOND simulations are shown in Fig.\ref{fig:masses}. A subtle point about the total masses of the two clusters is that we do not expect the mass to remain constant as we go back in time. Presumably they grew from a seed of negligible mass at high redshift (see discussion in Cameron \& Driver 2007). This tends to impede their freefall, reducing the maximum collision velocity to $\sim2900\kms$ by the estimate of Farrar \& Rosen (2007). To include this, without the impedance, we use the procedure of Wechsler et al. (2002) who used the relation 
\beq
M(z)=M(z=0.3)e^{-\alpha (z-0.3)}
\eeq
where $\alpha$ obviously encodes the speed of the accretion or assembly of the halo. Typical values used in their work are 0.5$< \alpha<$2.0. In Fig.\ref{fig:mz} we plot the mass enclosed within $r_{200}$ for the two clusters as functions of reshift for $\alpha$=0.5,1.0 and 1.5. Note we always keep a floor value of cluster mass of $m_{200}/50$ so the halo is never completely disassembled.

Another important point is whether mass integrated out beyond $r_{200}$ should be included, since the actual virial radius depends on both cosmology and redshift (Bullock et al.\ 2001). Indeed, the internal gravity of the main cluster has not reached $a_o$ by $r_{200}=2100kpc$ meaning the MOND dynamical mass has not yet saturated. However, recall that it takes $\sim (r_{200}-d)/v_{rel}=(2100kpc-400kpc)/3400\kms \sim 500Myr$ for the clusters to separate enough for this extra matter to even begin to manifest itself.  The cluster is also losing mass (backwards in time) due to accretion coupled with the fact there must be overdensities on the opposite side of the universe countering the influence of these overdensities. However, using Eq.\ref{eqn:nfwmass} it is straight-forward to include all the enclosed mass out to any radius because the parameters $r_{200}$ and $m_{200}$ do not explicitly force the enclosed mass to truncate at $r_{200}$, they simply define the shape of the profile. 

\section{The collision in MOND}
\protect\label{sec:mondbul}

In MOND, the basic modification of purely Newtonian dynamics is
\beq
\label{eqn:mond}
\mu(x) {\bf g} = {\bf g_N},
\eeq
where $g_N$ is the Newtonian acceleration computed in the usual way from the baryonic mass distribution, ${\bf g}$ is the actual acceleration (including the effective amplification due to MOND conventionally ascribed to DM), $a_o$ is the characteristic acceleration at which the modification becomes effective ($\sim 10^{-10}\;\textrm{m}\;\textrm{s}^{-2}$), $x = g/a_o$, and $\mu(x)$ is an interpolation function smoothly connecting the Newtonian and MOND regimes.  In the limit of large accelerations, $g \gg a_o$, $\mu \rightarrow 1$ and the Newtonian limit is obtained: everything behaves normally.  The MOND limit occurs only for exceedingly low accelerations, with $\mu \rightarrow x$ for $g \ll a_o$.  We implement two possible versions of the interpolation function:  the `standard' function traditionally used in fitting rotation curves:
\beq
\label{eqn:mum}
\mu={x \over \sqrt{1+x^2}}
\eeq
(e.g., Sanders \& McGaugh 2002), and the `simple' function found by Famaey \& Binney (2005) to provide a good fit the terminal velocity curve of the Galaxy:
\beq
\label{eqn:mub}
\mu={x \over 1+x}.
\eeq

A well known problem with implementing the MOND force law in numerical computations is that the original formulation (Eq \ref{eqn:mond}) does not conserve momentum (Felten 1984; Bekenstein 2007).  This was corrected with the introduction of a Lagrangian formulation of MOND (Bekenstein \& Milgrom 1984; Milgrom 1986) which has the modified Poisson equation
\beq
\label{eqn:aqual}
\grad \cdot [\mu(|\nabla \Phi|/a_o) \grad \Phi] = 4 \pi G \rho.
\eeq
This formulation has been shown to obey the necessary conservation laws (Bekenstein \& Milgrom 1984; Bekenstein 2007).  With some rearrangement, it leads to 
\beq
\label{eqn:curl}
\mu(x) {\bf g} = {\bf g_N} + \nabla \times {\bf h},
\eeq
which we recognize as Eq \ref{eqn:mond} with the addition of a curl field.

Unfortunately, implementing a numerical formulation of the modified Poisson equation is not a simple one-line change to typical N-body codes: this fails to obey the conservation laws. Instead, one needs an entirely different numerical approach than is commonly employed. Progress has been made along these lines (e.g., Brada \& Milgrom 1995, 1999; Ciotti, Londrillo, \& Nipoti 2006; Nipoti, Londrillo \& Ciotti 2007ab; Tiret \& Combes 2007; see also Nusser 2002; Knebe \& Gibson 2004), but we do not seek here a full N-body treatment of complex systems.  Rather, we wish to develop and apply a simple tool (Angus \& McGaugh, in preparation) that can provide some physical insight into basic problems. For the specific case of  the large collision velocity of the bullet cluster, it suffices to treat the curl field as a small correction to the center of mass motion (Milgrom 1986).  The external field effect (see Milgrom 1983a, Bekenstein 2007) is crudely approximated as a constant of appropriate magnitude (McGaugh 2004).  We checked the effect of varying the external field, which is modest. It is not possible to do better without complete knowledge of the mass distribution in the environment of the clusters.

When modeling the bullet cluster in MOND, Angus et al.\ (2007a) fitted the convergence map of Clowe et al. (2006) using spherical potential models for the four mass components. Their best fit gives masses for all four components in MOND and standard gravity. Unfortunately, the map is only sensitive out to 250kpc from the respective centres which neglects an over large portion of the dynamical mass. So, in order to remain consistent with the CDM simulations, we take the NFW profile and calculate what the MOND dynamical masses for the two commonly used interpolating functions (Eq.\ref{eqn:mum} \& \ref{eqn:mub}) are, as shown in Fig.\ref{fig:masses}. The Newtonian mass for the main cluster is twice that of the MOND dynamical mass with the standard $\mu$ and three times when the simple $\mu$ is used.

\begin{figure*}
\def\subfigtopskip{0pt}    %{4pt}
\def\subfigbottomskip{4pt}
\def\subfigcapskip{1pt}
\centering
\begin{tabular}{cc}

\subfigure{\label{fig:vcmd}
\includegraphics[angle=0,width=8.0cm]{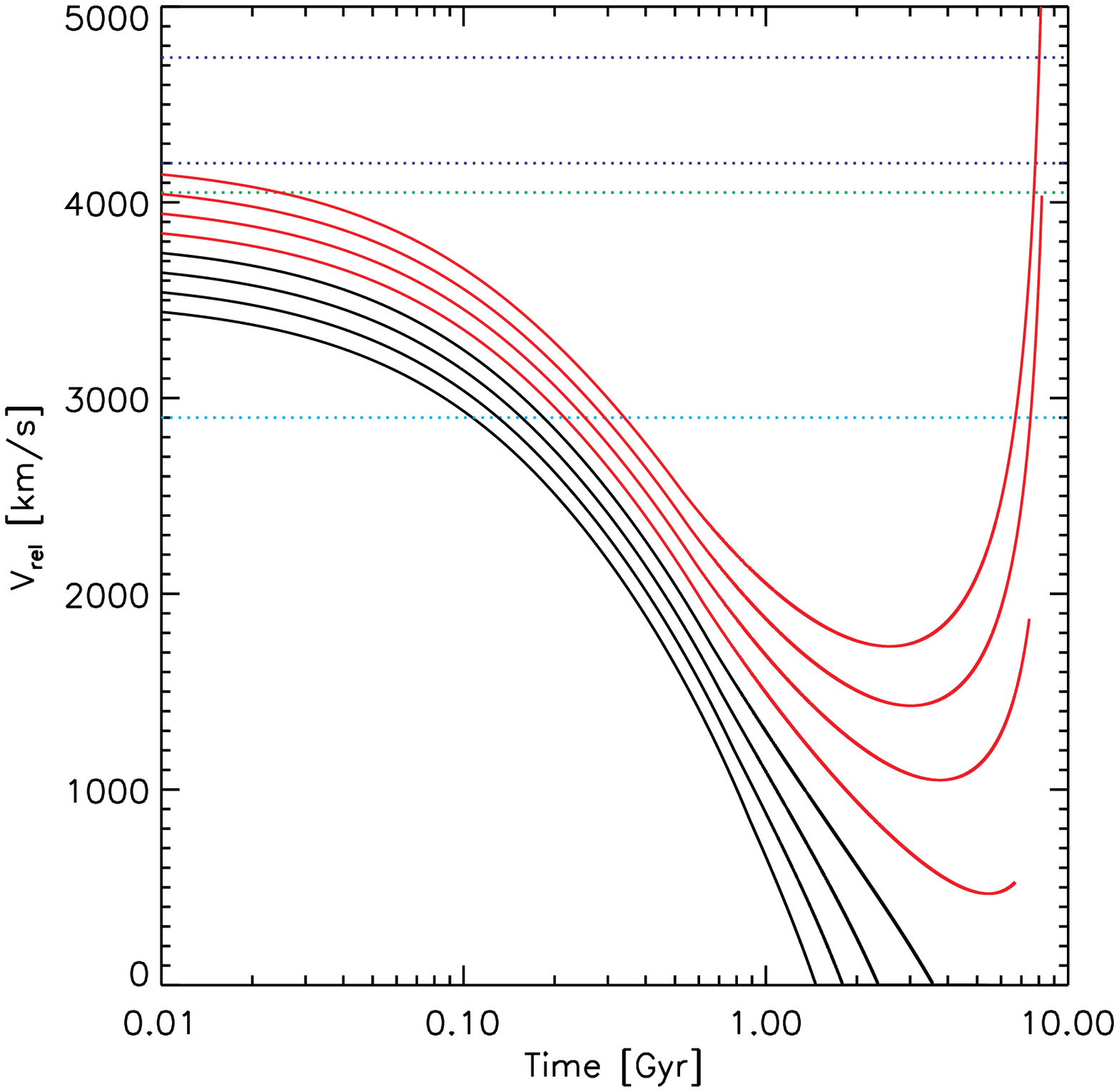}
}
&
\subfigure{\label{fig:vmond}
\includegraphics[angle=0,width=8.0cm]{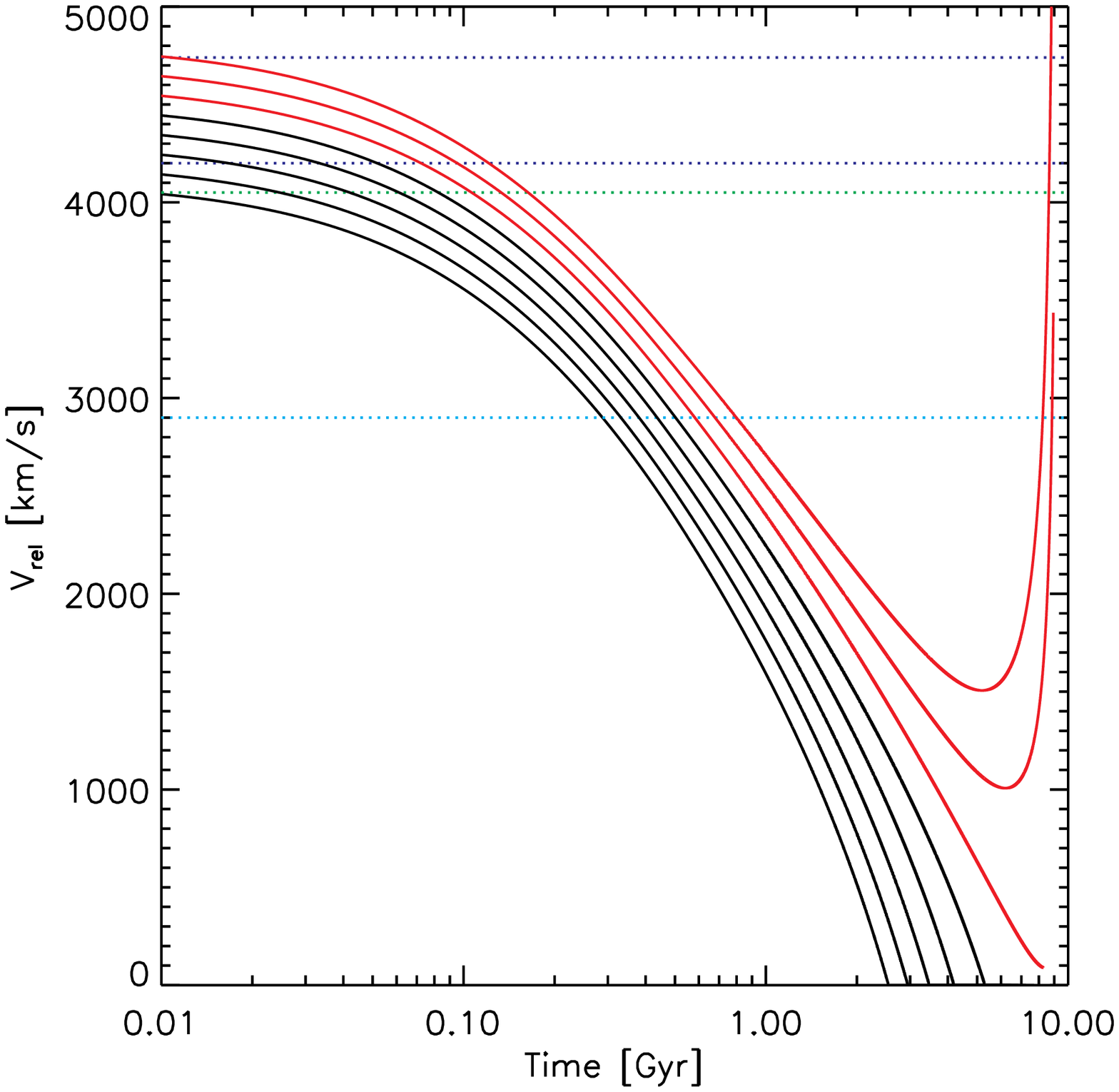}
}\\
\end{tabular}
\caption{Shows the relative velocity of the two clusters plotted against time (a) CDM and (b) MOND. Time=0Myr is the current (z=0.3) relative velocity of the two clusters with larger times corresponding to higher redshifts. Black lines correspond to relative velocities that are achievable, whereas red lines are not. In (a) we use the simulation (CDM2c) for which $\alpha$=1.0, d=425kpc and we truncate the halos at $r_{200}$. The relative velocities used are $v_{rel}$=3500-4200$\kms$ in intervals of 100$\kms$. In (b) we use the simulation (MONDst2) which uses the standard $\mu$ function and $\alpha$=0.5, d=425kpc. The relative velocities used are $v_{rel}$=4100-4800$\kms$ in intervals of 100$\kms$. The 4 dashed lines are the predicted relative velocities according to the mean and 1$\sigma$ error of the original relative velocity from Markevitch \& Vikhlinin (2007) in blue, the simulations of Milosavljevic (2007) in green and Springel \& Farrar (2007) in turquoise. The high observed collision velocity is more readily obtained in MOND than CDM.}
\protect\label{fig:vrel}
\end{figure*}

The mutual gravity imposed upon the sub cluster by the main cluster is 
\beq
\label{eqn:bulmond}
 \mu \left({|g_{sub}+g_{ex}| \over a_o}\right)g_{sub}=g_{n,sub}=-GM_{main}(d)/d^2
\eeq
and we simply swap the subscripts around to find the mutual gravity of the sub cluster upon the main cluster. Following on from above, $d$ is the distance between the two centres of mass and $M_{main}(d)$ is the enclosed mass within a radius $d$ from the centre of mass of the main cluster. The $g_{ex}$ is the external field limiting the MOND correction which comes from large scale structure and is always assumed orthogonal to the direction of $g_{sub}$, making the argument of the $\mu$ function more easily expressed as ${\left(g_{sub}^2+g_{ex}^2\right)^{1/2} \over a_o}$. The direction and amplitude of $g_{ex}$ is unknown at all times. The MONDian additional acceleration becomes minor when the acceleration drops below $g_{ex}$. We use $g_{ex} = a_o/30$ (Aguirre et al.\ 2001; McGaugh 2004) which is roughly the external field imposed on the Milky Way by M31 and vice versa (Famaey et al. 2007a, Wu et al. 2007). 

\section{N-body collision}

Our first attempt at simulating the collision in Newtonian gravity was using a standard N-body tree code. The benefit it gives is that in principal, we can more accurately compute the mutual gravity at the beginning of the simulation when the two clusters overlap. However, this is fraught with difficulties and inconsistencies. The first being that tidal effects undoubtedly stretch the two clusters and 2-body interactions may eject particles from the two halos. Therefore, it makes better sense to begin such a simulation from high redshift where the clusters are greatly separated and tidal effects are negligible and let them freefall in the expanding universe and when they collide, the tidal effects will be well accounted for. Of course, the problem is that it is not trivial to then sample collision velocities because the separation and time at which the two clusters began their freefall is not simply related. Moreover, the truncation of the two halos and different mass models are not easily varied. Nevertheless, we did attempt a CDM N-body model with truncation at $r_{200}$ for both halos. We found a similar result to that from the semi-analytical models of $3800 \kms$.  

\section{Results}

The ability of the two clusters that comprise the bullet to bring each other to a halt at a finite time in the past is sensitive to both the flavour of gravity at work and the true relative velocity. For velocities larger than the maximum, the relative velocity never reaches zero and increase sharply at early times (large z). The two clusters do not gravitate strongly enough to generate such high velocities and would have to have had a huge relative velocity towards each other in the early universe in order to overcome the Hubble expansion and fall together with such a high relative velocity at z=0.3. Fig.\ref{fig:vrel} shows how the relative velocity of the two clusters varies with time for a large sample of initial (meaning collisional) relative velocities for a CDM and MOND sample simulation. A difference of just 100$\kms$ can have a significant impact on the time required to generate such a large velocity and by the same token, the longer the two clusters free-fall, the larger a velocity they can generate. Sadly, there is only a finite time ($\sim$ 9Gyr)) since the Big Bang for this to happen.

\begin{table*} 
\begin{tabular}{|c|c|c|c|c|} 
${\rm Model}$ &Max $V_{rel}$ [$\kms]$ & Truncation Radius &$\alpha$&Gravity\\
 CDM1a & 4500 & $r_1$ & 0.0& Newtonian\\
 CDM1b & 4200 & $r_1$ & 1.0& Newtonian\\
 CDM2a & 4000 & $r_{200}$ & 0.0& Newtonian\\
 CDM2b & 3900 & $r_{200}$ & 0.5& Newtonian\\
 CDM2c & 3800 & $r_{200}$ & 1.0& Newtonian\\
 CDM2d & 3800 & $r_{200}$ & 1.5& Newtonian\\
 MONDst1 & 4800 & $r_{200}$ & 0.0& MOND-standard $\mu$\\
  MONDst2 & 4500 & $r_{200}$ & 0.5& MOND-standard $\mu$\\
MONDsi1 & 4600 & $r_{200}$ & 0.0& MOND-simple $\mu$\\
MONDsi2 & 4500 & $r_{200}$ & 0.5& MOND-simple $\mu$\\
\end{tabular} 

\medskip 
\caption{Shows the parameters used in the different models and gives the maximum attainable relative velocity for each.} 
\end{table*} 

In Table 1 we've put the key results of the simulations so as to give the reader a feel for what the maximum relative velocity that can be achieved is. Each velocity is that achieved with an initial separation of 425kpc, where taking 350 or 500kpc induces an increase or decrease of 100$\kms$ which we take as the minimum error. The most extreme CDM model is to have no truncation of the DM halos, extending them out to $r_1$. This absurd extreme allows a maximum relative velocity of 4500$\kms$. Then, if we still allow the halos to extend to $r_1$, but account for some assembly of the halos with $\alpha=1$ then the relative velocity reduces to 4200$\kms$.

More realistically, if we truncate the halos at $r_{200}$ and try four different halo assembly rates such that $\alpha=$0.0, 0.5, 1.0 \& 1.5 we get respective maximum relative velocities of 4000, 3900, 3800 and 3800$\kms$. These numbers represent the plausible maximum relative velocities in the CDM framework. 

For the MOND case we ran simulations with both the simple (Eq.\ref{eqn:mub}) and standard (Eq.\ref{eqn:mub}) $\mu$ functions. The standard function leads to higher dynamical masses from the NFW profile, but lower 2-body gravity. The standard (simple) function with no accretion and with $\alpha=0.5$ generate 4800 (4600) and 4600 (4500)$\kms$ respectively and for comparison, the maximum CDM velocity with those reduced masses is just 2700 (2300)$\kms$. This is a clear demonstration of the expectation in MOND for larger peculiar velocities. We use the lower assembly parameter $\alpha=0.5$ because structure is expected to form more swiftly in MOND (Sanders 1998, 2001). 

An important factor is that of the fitted NFW density profile to the convergence map, in which matter is extrapolated to 2100kpc and 1000kpc for the main and sub cluster respectively. Presumably the significance of the detection of this mass is negligible and the NFW fit has been made assuming if we know the details in the central 250kpc, then we know the density out to $r_{200}$. The mass sheet degeneracy is broken by constraining the mass at the edges of the fit based on the slope of the profile in the inner regions - but if the mass profile is wrong then it could lead to the completely wrong measurement for the value of the mass sheet (Clowe, de Lucia and King 2004). 

All of this means that the density profiles of the two clusters could be moderately different in reality. However, the actual shape of any profile is less important to the relative velocity than simply the normalisation of the total mass. To this end we have simulated the collision with 10\% more and 10\% less mass for both clusters (with assembly parameter $\alpha=1$). The effect is to increase (10\% more mass) or decrease (10\% less mass) the relative velocity by 200$\kms$ from 4800$\kms$ for model MONDst1.

Another concern is that the clusters are unlikely to be spherically symmetric (Buote \& Canizares 1996) and are presumably elongated in the direction of motion. Again this could lead to an incorrect density profile, whereas ellipticity itself would have little effect on our results.

\section{Summary}

We have constructed specific mass models for the bullet cluster in both CDM and MOND. We integrate backwards from the observed conditions to check whether the large ($\sim 4700\kms$) apparent transverse velocity can be attained in either context.  We find that it is difficult to achieve $v_{rel} > 4500\kms$ under any conditions. Nevertheless, within the range of the uncertainties, the appropriate velocity occurs fairly naturally in MOND. In contrast, $\Lambda$CDM models can at most attain $\sim 3800\kms$ and are more comfortable with considerably smaller velocities.

Taken at face value, a collision velocity of $4700\kms$ constitutes a direct contradiction to $\Lambda$CDM. Ironically, this cluster, widely advertised as a fatal observation to MOND because of the residual mass discrepancy it shows, seems to pose a comparably serious problem for $\Lambda$CDM. It has often been the case that observations which are claimed to falsify MOND turn out to make no more sense in terms of dark matter.

Two critical outstanding issues remain to be clarified. The first is the exact density profiles and virial masses of the two clusters and the second is how the observed shock velocity relates to the actual collision velocity of the two gravitating masses. 
The recent simulations of Springel \& Farrar (2007) and Milosavljevic et al. (2007) seem to suggest that, contrary to naive expectations, hydrodynamic effects reduce the relative velocity of the mass with respect to the shock. A combination of effects is responsible, being just barely sufficient to reconcile the data with $\Lambda$CDM.  Hydrodynamical simulations are notoriously difficult, and indeed these two recent ones do not agree in detail.  It would be excellent to see a fully self-consistent simulation including both hydrodynamical effects and a proper mass model and orbital computation like that presented here.

There are a number of puzzling aspects to the hydrodynamical simulations. First of all, Springel \& Farrar (2007) use Hernquist profiles for the DM distribution in the clusters and not NFW halos. Furthermore, they find that the morphology of the bullet is reproduced only for a remarkably dead head-on collision.  If the impact parameter is even 12kpc --- a target smaller than the diameter of the Milky Way --- quite noticeable morphological differences ensue. This can be avoided if the separation of mass centres happens to be along our line of sight --- quite a coincidence in a system already remarkable for having the vector of its collision velocity almost entirely in the plane of the sky. Furthermore, the mass models require significant tweaking from that infered from the convergence map and are unable to reproduce the currently observed, post merger positions of the gas and DM. It appears to us that only the first rather than the last chapter has been written on this subject.  Getting this right is of the utmost importance, as the validity of both paradigms rests on the edge of a knife, separated by just a few hundred $\kms$.

More generally, the frequency of bullet-like clusters may provide an additional test.  The probability of high collision velocities drops with dramatic rapidity in $\Lambda$CDM (Hyashi \& White 2006).  In contrast, somewhat higher velocities seem natural to MOND.  Naively it would seem that high impact velocity systems like the bullet would be part and parcel of what might be expected of a MOND universe. With this in mind, it is quite intriguing that many bullet cluster like systems have been detected (although none quite as unique). The dark ring around Cl0024+17 tentatively observed by Jee et al. (2007; see also Famaey et al. 2007c), the dark core created by the ``train wreck" in Abell 520 by Mahdavi et al. (2007), Cl0152+1357 (Jee et al. 2005a), MS1054+0321 (Jee et al. 2005b) and the line of sight merger with $>3000\kms$ relative velocity observed by Dupke et al. (2007) for Abell 576 may all provide examples and potential tests.

\section*{Acknowledgements}
We acknowledge discussions with Benoit Famaey, Tom Zlosnik, Douglas Clowe, HongSheng Zhao, Greg Bothun, Moti Milgrom, Bob Sanders and Ewan Cameron. GWA thanks Steve Vine for his N-body tree code. GWA is supported by a PPARC scholarship.  The work of SSM is supported in part by NSF grant AST0505956.
%John Reid, Iain Mackenzie, Barry Gibson and PJ Bruce.

\label{lastpage}

\end{document}